\documentclass[preprint,12pt]{elsarticle} 
\usepackage{graphicx} 
\usepackage{amssymb} 

\journal{Nuclear Physics A} 

\begin{document}

\begin{frontmatter} 

\title{Kaonic atoms and in-medium $K^-N$ amplitudes} 

\author[a]{E.~Friedman}
\author[a]{A.~Gal}  
\address[a]{Racah Institute of Physics, The Hebrew University, 
Jerusalem 91904, Israel} 

\date{\today}

\begin{abstract}

Recent work on the connection between in-medium subthreshold $K^-N$ amplitudes 
and kaonic atom potentials is updated by using a next to leading order 
chirally motivated coupled channel separable interaction model that reproduces 
$\bar KN$ observables at low energies, including the very recent SIDDHARTA 
results for the atomic $K^-$-hydrogen $1s$ level shift and width. 
The corresponding $K^-$-nucleus potential is evaluated self-consistently 
within a single-nucleon approach and is critically reviewed with respect 
to empirical features of phenomenological optical potentials. The need to 
supplement the single-nucleon based approach with multi-nucleon interactions 
is demonstrated by showing that additional empirical absorptive and dispersive 
terms, beyond the reach of chirally motivated $K^-$-nucleus potentials, are 
required in order to achieve good agreement with the bulk of the data on 
kaonic atoms. 

\end{abstract}


\end{frontmatter} 

\section{Introduction}
\label{sec:intro}

The bulk of the data on strong interaction effects in kaonic atoms is due to 
experiments of over three decades ago. With the exception of the very light 
atoms of $K^-$H and $K^-$$^4$He the rest of the data could be described rather 
well with the help of $K^-$-nucleus optical potentials \cite{FG07,BFG97}. 
Recent experiments on $K^-$H and $K^-$$^4$He with much reduced background 
removed the `puzzles' with these two atoms \cite{Iwa97,IHN98,Baz11,Oka07}.  
However, the depth of the attractive $K^-$-nucleus real potential, which in 
phenomenological analyses came out in the range of 150-200 MeV \cite{FGB93}, 
presented a theoretical challenge in as much as in-medium chiral 
{\it threshold} $K^-N$ scattering amplitude input led to a lower value of 
order 120 MeV \cite{WKW96}, or even to a considerably lower value of 40-50 MeV 
\cite{ROs00}. This outstanding discrepancy is of current interest, since it is 
relevant to the role of $K^-$ mesons in multistrange self-bound matter and in 
compact stars \cite{GFGM07,SSS08}. The problem has been largely resolved very 
recently \cite{CFG11,CFG11a} noting that in-medium chiral {\it subthreshold} 
$K^-N$ scattering amplitudes provide the relevant input and thereby 
demonstrating the need to supplement the model by multi-nucleon terms, 
as discussed in the present work. This leads to deep real potentials in 
agreement with the purely phenomenological analyses. 

The present paper is an update of Refs.~\cite{CFG11,CFG11a}, based on a recent 
in-medium coupled channel chirally motivated separable interaction model which 
produces good fits to all the low energy antikaon-nucleon data, including the 
latest $K^-$H atom results from the SIDDHARTA experiment \cite{Baz11}. 
Section \ref{sec:KNampl} outlines the self-consistent handling of subthreshold 
$K^-N$ amplitudes while section \ref{sec:Knucleus} deals with the resulting 
$K^-$-nucleus amplitudes. Section \ref{sec:potls} reports on global optical 
model fits to kaonic atom data and the last section provides summary and 
conclusions.

\section{$K^-N$ scattering amplitudes}
\label{sec:KNampl}

The potential experienced by a $K^-$ meson of energy $E_K^{\rm lab}=\omega_K$ 
interacting with a nucleus of density $\rho$ is given in the single-nucleon 
approximation by 
\begin{equation} 
V_{K^-}(\omega_K;\rho)=-\:\frac{2\pi}{\omega_K}\:(1+\frac{\omega_K}{m_N})\:
F_{K^-N}(\vec p,\sqrt{s};\rho)\:\rho, 
\label{eq:t} 
\end{equation} 
where $F_{K^-N}(\vec p,\sqrt{s};\rho)$ is the in-medium $K^-N$ scattering 
amplitude, reducing in the low-density limit $\rho\to 0$ to the free-space 
$K^-N$ c.m. forward scattering amplitude $F_{K^-N}(\vec p,\sqrt{s})$, $\vec p$ 
is the relative $K^-N$ momentum, $s=(E_K+E_N)^2-({\vec p}_K+{\vec p}_N)^2$ is 
the Lorentz invariant Mandelstam variable equal to the square of the total 
$K^-N$ energy in the two-body c.m. frame, and the nucleon energy $E_N$ is 
approximated by its mass $m_N$ in the kinematical factor in front of 
$F_{K^-N}$. The in-medium amplitude $F_{K^-N}(\vec p,\sqrt{s};\rho)$ in this 
work is a chirally motivated amplitude constructed within a full octet $0^-$ 
meson--octet $1/2^+$ baryon coupled channel separable interaction model 
\cite{CSm10,CSm11} which in its latest next to leading order (NLO) version 
NLO30 \cite{CSm11} incorporates the recent SIDDHARTA data for the atomic 
$K^-$H $1s$ level shift and width \cite{Baz11}. In this separable interaction 
model, the in-medium coupled channel scattering amplitudes assume the form 
\begin{equation} 
F_{ij}(p,p',\sqrt{s};\rho)=g_{i}(p)f_{ij}(\sqrt{s};\rho)g_{j}(p'), 
\label{eq:sep} 
\end{equation} 
with form factors $g_{j}(p)=\alpha_{j}^2/(p^2+\alpha_{j}^2)$. The momentum 
dependence introduced by the form factor $g_{K^-N}$ in the separable 
interaction model of Refs.~\cite{CSm10,CSm11} is relatively weak for the 
applications discussed in the present work and is secondary to the strong 
energy dependence of the reduced amplitude $f_{K^-N}$ generated by the 
$\Lambda(1405)$ subthreshold resonance. 

\begin{figure}[htb] 
\begin{center} 
\includegraphics[width=0.6\textwidth]{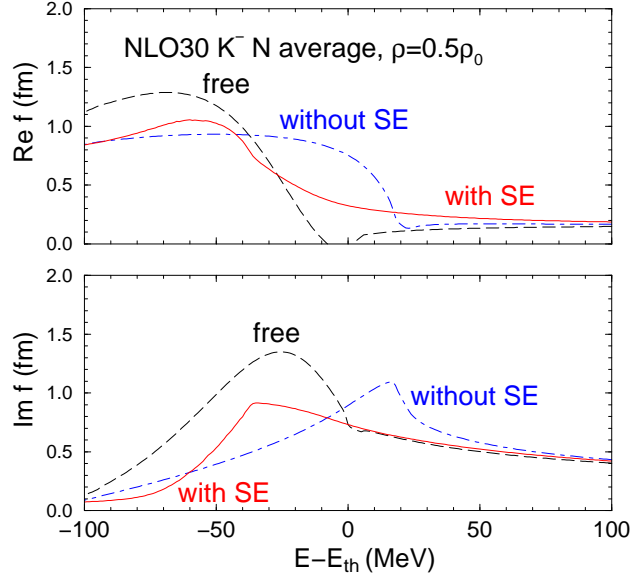} 
\caption{Energy dependence of the c.m. $K^-N$ reduced amplitude 
(\ref{eq:isoscalar}) in version NLO30 of the chiral model \cite{CSm11} below 
and above $E_{\rm th}=m_K+m_N=1432$~MeV. Dashed curves: free-space amplitude; 
dot-dashed curves: Pauli blocked amplitude at $0.5\rho_0$; solid curves: 
including meson and baryon self energies (SE), also at $0.5\rho_0$.} 
\label{fig:NLO30} 
\end{center} 
\end{figure} 

Free-space and in-medium reduced amplitudes for half nuclear matter 
density $\rho=0.5\rho_0$ are shown in Fig.~\ref{fig:NLO30} for the 
isospin-averaged combination 
\begin{equation} 
f_{K^-N}=\frac{1}{2}(f_{K^-p}+f_{K^-n})=\frac{3}{4}f_{I=1}+\frac{1}{4}f_{I=0}, 
\label{eq:isoscalar} 
\end{equation} 
corresponding to symmetric nuclear matter. Similar results are obtained at 
full nuclear matter density $\rho_0=0.17~{\rm fm}^{-3}$. Of the two in-medium 
amplitudes shown in the figure, the one marked ``without SE" imposes Pauli 
blocking on intermediate $\bar K N$ states for $\rho \neq 0$ \cite{WKW96}, 
whereas the one marked ``with SE" adds self consistently hadron self energies 
in intermediate states \cite{CFG01}, following a procedure suggested in 
Ref.~\cite{Lut98}. The real part of all three amplitudes exhibits strong 
energy dependence, switching from weak attraction above $K^-N$ threshold to 
strong attraction below threshold. As a rule of thumb, ${\rm Re}~f=1$ fm 
translates into a sizable attraction ${\rm Re}~V_{K^-}\approx -100$ MeV. 
The imaginary part of these amplitudes exhibits a peak, related to the 
subthreshold $\Lambda(1405)$ resonance, with a steep decrease at lower 
energies, becoming vanishingly small near the $\pi\Sigma$ threshold about 
100 MeV below the $K^-N$ threshold. In between the two limits of the energy 
scale in Fig.~\ref{fig:NLO30}, $E-E_{\rm th}=\pm 100$ MeV, 
the three amplitudes differ appreciably from each other. At threshold, 
in particular, the real part of the ``with SE" amplitude is about half of 
that ``without SE", corresponding to a depth $-{\rm Re}~V_{K^-}(\rho_0)\approx 
40\!-\!50$~MeV, in agreement with Ramos and Oset \cite{ROs00}. 

\begin{figure}[htb] 
\begin{center} 
\includegraphics[width=0.6\textwidth]{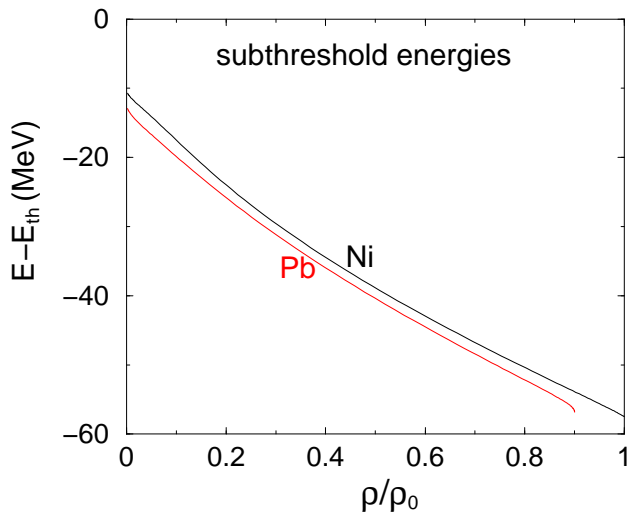} 
\caption{Subthreshold energies as function of nuclear density, see text.} 
\label{fig:NLO30Evsrho} 
\end{center} 
\end{figure}

It was recognized in the early 1970s that the strong energy dependence 
of the two-body amplitude, particularly in the subthreshold region where 
the $\bar K N$ quasibound state $\Lambda(1405)$ dominates, provides the 
underlying structure for $K^-$ nuclear interactions at and near the $K^-$ 
nucleus threshold \cite{Wyc71,BTo72,Roo75}. This idea has been reformulated 
and applied recently in Refs.~\cite{CFG11,CFG11a} to a comprehensive study of 
kaonic atoms. The essential idea is to replace the two-body variables $\vec p$ 
and $\sqrt{s}$ of the in-medium scattering amplitude by appropriate density 
dependent averages in the nuclear medium. This may be summarized by the 
following relationships: 
\begin{equation} 
\sqrt{s} \rightarrow E_{\rm th} - B_N - B_K - \xi_N\frac{p_N^2}{2m_N} - 
\xi_K\frac{p_K^2}{2m_K}, 
\label{eq:sqrts}  
\end{equation} 
upon neglecting quadratic terms in the binding energies $B_K=m_K-E_K, 
B_N=m_N-E_N$ near threshold ($E_{\rm th}=m_N+m_K$), and 
\begin{equation} 
p^2, ~{p^\prime}^2 \rightarrow \xi_N\xi_K(2m_K\frac{p_N^2}{2m_N}+
2m_N\frac{p_K^2}{2m_K}),  
\label{eq:p^2} 
\end{equation} 
where $\xi_{N(K)}=m_{N(K)}/(m_N+m_K)$ in both of these substitutions. 
Replacing in Eqs.~(\ref{eq:sqrts}) and (\ref{eq:p^2}) the kinetic energy 
$p_K^2/(2m_K)$ in the local density approximation by 
$-B_K -{\rm Re}\:{\cal V}_{K^-}(\rho)$ where ${\cal V}_{K^-}=V_{K^-}+V_c$, 
with $V_c$ the $K^-$ Coulomb potential generated by the finite-size nuclear 
charge distribution, and approximating the nucleon kinetic energy 
$p_N^2/(2m_N)$ in the Fermi gas model by $23\,(\rho/\rho_0)^{2/3}$~MeV, 
Eqs.~(\ref{eq:sqrts}) and (\ref{eq:p^2}) become 
\begin{equation} 
{\sqrt{s}} \approx E_{\rm th} - B_N - \xi_N B_K - 
15.1(\frac{\rho}{\rho_0})^{2/3}+\xi_K{\rm Re}\:{\cal V}_{K^-}(\rho), \;\;\; 
(\rm {in~MeV})   
\label{eq:sfinal} 
\end{equation} 
where all the terms following $E_{\rm th}$ on the r.h.s. are negative, 
thus implementing the anticipated downward energy shift into the $K^-N$ 
subthreshold energy region, and 
\begin{equation} 
p^2 \approx \xi_N\xi_K[2m_K 23(\rho/\rho_0)^{2/3}-
2m_N(B_K+{\rm Re}\:{\cal V}_{K^-}(\rho))], \;\;\; (\rm {in~MeV}) 
\label{eq:p^2final} 
\end{equation} 
where both terms on the r.h.s. are positive for attractive potentials 
$V_{K^-}$. The dominant contribution in Eqs.~(\ref{eq:sfinal}) and 
(\ref{eq:p^2final}) arises from ${\rm Re}\:{\cal V}_{K^-}(\rho)$, 
resulting in a downward energy shift of up to 60 MeV as demonstrated in 
Fig.~\ref{fig:NLO30Evsrho}, and in values of $p(\rho_0),p^\prime(\rho_0)$ 
as high as 275 MeV/c for $K^-$ nuclear potential depths reaching 180 MeV 
in phenomenological studies \cite{FG07}. These momenta are well within 
the NLO30 momentum dependence scale $\alpha_{\bar K N}=700$ MeV/c. 

Having transformed the dependence of the in-medium scattering amplitudes 
$F_{K^-j}(\vec p,\sqrt{s},\rho)$ ($j=p,n,N$) on $\vec p$ and $\sqrt{s}$ 
into a density dependence, we denote the resultant in-medium scattering 
amplitudes by ${\cal F}_{K^-j}(\rho)$. In order to allow for different 
proton and neutron distributions in the actual calculations detailed below, 
the in-medium amplitude ${\cal F}_{K^-N}(\rho)$ which should substitute for 
$F_{K^-N}(\vec p,\sqrt{s},\rho)$ in the construction of the $K^-$-nucleus 
potential Eq.~(\ref{eq:t}) is further replaced by an effective in-medium 
amplitude ${\cal F}_{K^-N}^{\rm eff}(\rho)$: 
\begin{equation} 
\label{eq:effA} 
{\cal F}_{K^-N}^{\rm eff}(\rho)\rho(r) = 
{\cal F}_{K^-p}(\rho)\rho_p(r) + {\cal F}_{K^-n}(\rho)\rho_n(r), 
\end{equation} 
with $\rho_p$ and $\rho_n$ normalized to $Z$ and $N$, respectively, and 
$Z+N=A$. The reduced 
amplitudes $f_{K^-p}$ and $f_{K^-n}$ are evaluated at $\sqrt{s}$ given by 
Eq.~(\ref{eq:sfinal}), where the $K^-$ atomic binding energy $B_K$ is 
neglected with respect to the average nucleon binding energy $B_N\approx 8.5$ 
MeV.{\footnote{The precise value used for $B_N$ in our kaonic atom global fit 
hardly matters within reasonable limits. Sensitivity to $B_N$ is expected in 
studies limited to light nuclei, as exhibited recently by analyzing FINUDA 
data \cite{Fin11} of $\Lambda$ hypernuclear formation with stopped $K^-$ 
mesons on targets from lithium to oxygen \cite{CFGK11}.}} A similar 
approximation is made in Eq.~(\ref{eq:p^2final}) for $p^2$ when using the 
form factors $g_{K^-N}(p)$ of Eq.~(\ref{eq:sep}). The $K^-$-nucleus potential 
$V_{K^-}(\rho)$ is calculated by requiring self consistency in solving 
Eq.~(\ref{eq:sfinal}) with respect to ${\rm Re}\:V_{K^-}$, i.e., the value 
of ${\rm Re}\:V_{K^-}(\rho)$ in the expression for $\sqrt s$ and in the form 
factors $g_{K^-N}$ has to agree with the resulting ${\rm Re}\:V_{K^-}(\rho)$. 
This is done at each radial point and for every target nucleus in the data 
base.

\section{$K^-$-nucleus scattering amplitudes} 
\label{sec:Knucleus} 

The present model transforms the energy dependence of subthreshold effective 
amplitudes, Eq.~(\ref{eq:effA}), into density dependence. This transformation 
is hardly sensitive to the nucleus involved, as is seen in 
Fig.~\ref{fig:NLO30Evsrho} calculated in the ``without SE" version of the 
NLO30 model. In the ``with SE" version the energies for a given density 
differ from the plotted values by 2-3 MeV. The energy shifts do not vanish 
for zero density because we have used a fixed average nucleon binding energy 
of 8.5 MeV. Replacing it by a position-dependent $B_N \rightarrow B_N \rho(r)/
\bar\rho$, in order to satisfy the low-density limit, causes the energy shift 
to vanish far outside the nucleus, with minor overall effects on the present 
results. It is seen from the figure that, e.g., for a  density of 50\% of 
nuclear matter density, the downward energy shift is $\approx 40$~MeV which, 
from Fig.~\ref{fig:NLO30}, implies a real amplitude at least twice larger 
than the threshold value in the ``with SE" version. 

The density dependencies of the real and of the imaginary part of the 
effective amplitude ${\cal F}_{K^-N}^{\rm eff}$ are of particular interest 
because they are related to characteristic features of phenomenological 
optical potentials. It was shown already in 1993 \cite{FGB93} that with 
empirical density-dependent potentials, where the effective $K^-N$ scattering 
amplitude within a $t\rho$ model depends on the density, improved fits to the 
data were obtained compared to fits using fixed amplitudes. It was observed 
that in addition to the increased depth of the best-fit real potentials, 
these were characterised by {\it compression} relative to the corresponding 
nuclear densities, with r.m.s. radii of the real potential smaller than the 
corresponding nuclear radii. Reduced r.m.s. radii of optical potentials mean 
that the underlying in-medium $K^-N$ interaction increases with density, 
a robust feature that is insensitive to details. 

\begin{figure}[thb] 
\begin{center} 
\includegraphics[width=0.6\textwidth]{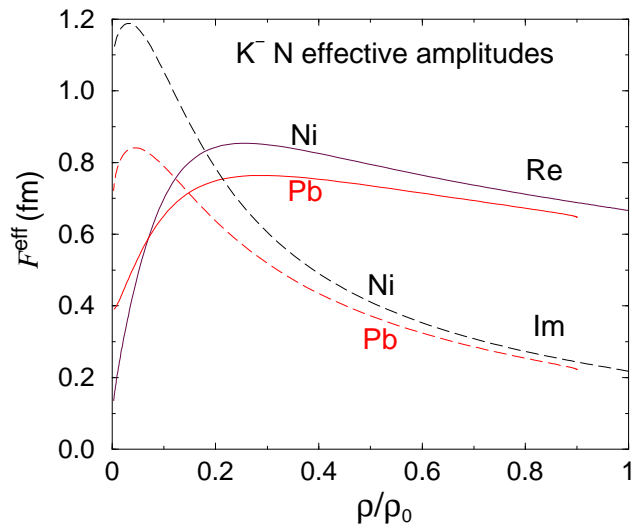} 
\caption{Effective amplitudes as function of nuclear density in model NLO30 
without SE.} 
\label{fig:effA} 
\end{center} 
\end{figure} 

Figure \ref{fig:effA} shows the NLO30 effective amplitudes for Ni and Pb 
as function of nuclear density, calculated in the ``without SE" version. 
Qualitatively similar results are obtained also in the ``with SE" version, 
see Ref.~\cite{CFG11}. Regions of low density, i.e. large radii, are the 
most effective in determining the r.m.s. radius of a distribution and the 
sharp rise of the real part in the extreme surface region can lead to 
compression of the real potential. The opposite dependence is observed for 
the imaginary part, thus implying {\it inflation} of the imaginary potential 
relative to the nuclear density. Quantitatively, however, the change of 
r.m.s. radii relative to nuclear densities is found to disagree with 
the empirical trends of Ref.~\cite{FGB93}, with too little compression 
for the real part and far too strong inflation for the imaginary part. 
Inevitably this is reflected in the quality of agreement with the data, 
as demonstrated in the next section. 

Strong-interaction effects in kaonic atoms are dominated by absorption, as is 
evident from level widths being significantly larger than the corresponding 
level shifts. Moreover, the shifts are always repulsive although the real 
potential is attractive, again pointing to the dominance of absorption. 
Therefore it is argued that the above marked decrease of the imaginary part 
of the effective scattering amplitude is the main deficiency of the present 
model, a decrease originating in the sharp decrease of the imaginary part of 
the free amplitude (Fig.~\ref{fig:NLO30}) towards the $\pi\Sigma$ threshold, 
which is typical of the single-nucleon approach. Consequently it is reasonable 
to expect that additional, multi-nucleon terms are required to obtain good fit 
to the data.

\section{$K^-$-nucleus optical potentials} 
\label{sec:potls} 

Strong interaction level shifts and widths in kaonic atoms have been 
calculated by solving a Klein-Gordon equation \cite{FG07} with the optical 
potential of Eq.~(\ref{eq:t}) transformed to the $K^-$-nucleus c.m. system, 
and where the in-medium $K^-N$ scattering amplitude is given by the effective 
amplitude Eq.~(\ref{eq:effA}): 
\begin{equation} 
V_{K^-}=-\frac{2\pi}{\mu}(1+\frac{A-1}{A}\frac{\mu}{m_N})
{\cal F}_{K^-N}^{\rm eff}(\rho)\rho(r), 
\label{eq:Vopt} 
\end{equation} 
with $\mu$ the kaon-nucleus reduced mass and $\rho=\rho_p+\rho_n$. 
Two-parameter Fermi distributions were used for both densities, with $\rho_p$ 
obtained from the known charge distribution by unfolding the finite size of 
the charge of the proton. For $\rho_n$ averages of the `skin' and `halo' forms 
of Ref.~\cite{TJL01} were adopted with the difference between r.m.s. radii 
given by $r_n-r_p = (N-Z)/A-0.035$~fm.

\begin{figure}[thb] 
\begin{center} 
\includegraphics[width=0.6\textwidth]{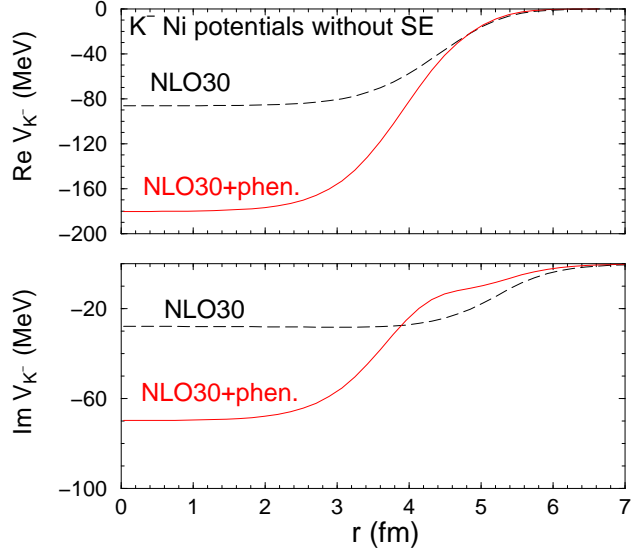}
\caption{$K^-$ nuclear potentials for $K^-$ atoms of Ni. 
Dashed curves: derived from in-medium NLO30 amplitudes; 
solid curves: plus phenomenological terms from global fits.} 
\label{fig:Nipotl} 
\end{center} 
\end{figure} 

\begin{figure}[thb] 
\begin{center} 
\includegraphics[width=0.6\textwidth]{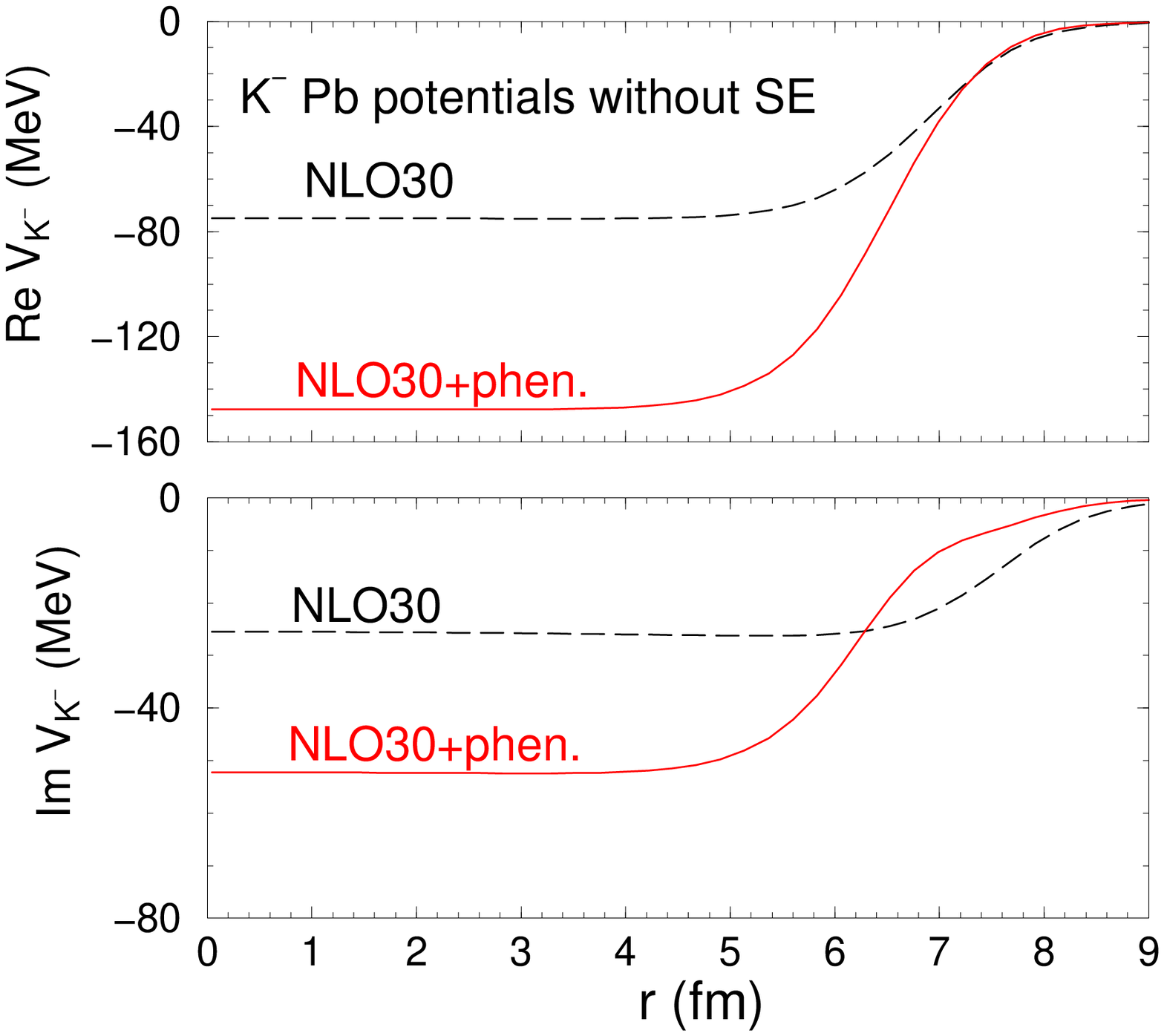} 
\caption{$K^-$ nuclear potentials for $K^-$ atoms of Pb. 
Dashed curves: derived from in-medium NLO30 amplitudes; 
solid curves: plus phenomenological terms from global fits.}  
\label{fig:Pbpotl} 
\end{center} 
\end{figure} 

Figures \ref{fig:Nipotl} and \ref{fig:Pbpotl} show optical potentials for 
$K^-$-Ni and for $K^-$-Pb, respectively. The potentials marked NLO30 follow 
directly from the in-medium $K^-N$ amplitudes in model NLO30, without any 
adjustable parameters. The agreement with the full data set of 65 points, 
covering the whole periodic table, is poor, with $\chi ^2$ per point in the 
range of $10-12$. This is not surprising in view of the obvious deficiency 
of the single-nucleon approach where the imaginary part of the amplitude 
goes down rapidly towards far subthreshold energies. Adding to the potential 
an empirical term linear in the nuclear density does not improve much the 
fit to kaonic atoms data and only by further addition of a $\rho ^2 /\rho_0$ 
or a $\rho (\rho /\rho_0)^2$ term good fits to the data are possible. 
Both imaginary and real parts of the additional phenomenological potential 
are then found to be dominated by  $\rho ^2$ or $\rho ^3$ terms which are 
likely to represent multi-nucleon absorptive and dispersive contributions, 
respectively. Figures~\ref{fig:Nipotl} and \ref{fig:Pbpotl} also show 
potentials obtained when adjustable $b\rho + B\rho^2 /\rho_0$ terms are 
added and the four parameters $b$ and $B$ are determined by requiring 
best fit to the data. The quality of the fits is then quite acceptable, 
with $\chi ^2$ per point of 2 to 2.3. Qualitatively similar results to 
those displayed in Figs.~\ref{fig:Nipotl} and \ref{fig:Pbpotl} are obtained 
also in the ``with SE" version, in agreement with the discussion for Ni in 
Ref.~\cite{CFG11}. 

Considering values of the potentials at the nuclear center, the additional 
phenomenological part appears too large to be regarded as a correction term 
to the basic NLO30 amplitude. However, values of the potential at the center 
are rather meaningless in the context of kaonic atom observables. The 
sensitivity of calculated level shifts and widths to the $K^-$ nuclear 
potentials was found \cite{BFr07} to be around the nuclear surface and 
certainly not at the center. With that in mind and focusing on the imaginary 
potential as noted above, it is remarkable that the additional term modifies 
the shape of the imaginary potentials in the surface region, bringing their 
r.m.s. radii closer to empirical values. Changes of the imaginary potentials 
near the surface due to the phenomenological term are of the order of 30\% of 
the NLO30 potentials, consistent with the fraction of multi-nucleon absorption 
estimated from experiments with emulsions and bubble chambers \cite{VVW77}. 
The emerging phenomenology is similar to that for $V_{\pi^-}$ in pionic 
atom studies where theoretically motivated single-nucleon contributions 
are supplemented by phenomenological $\rho^2$ terms representing $\pi NN$ 
processes \cite{EE66}. (See also Ref.~\cite{FG07}).

\section{Summary and conclusions} 
\label{sec:summ} 

A simple ansatz for transforming the strong energy dependence of subthreshold 
$K^-N$ scattering amplitudes in the nuclear medium to appropriate density 
dependent averages was presented and employed in global analyses of kaonic 
atom data, following Refs.~\cite{CFG11,CFG11a}. With chirally motivated 
coupled channel separable interaction scattering amplitudes in model NLO30 
\cite{CSm11} that respect the low energy ${\bar K}N$ data, including the 
recent SIDDHARTA results for kaonic hydrogen \cite{Baz11}, the connection 
between this model and deep real optical potentials was re-established. 
Effective $K^-$-nucleus amplitudes were derived self-consistently and were 
critically reviewed with respect to empirical features of phenomenological 
optical potentials. We focused in the present update on the in-medium effects 
arising exclusively from the strong energy dependence of subthreshold $K^-N$ 
amplitudes, using for this purpose the ``without SE" version of the NLO30 
in-medium model. The introduction of self-energy effects in the ``with SE" 
version is necessarily model dependent to some extent. Nevertheless, 
all of our findings and conclusions hold true in both ``without SE" and 
``with SE" versions, with minor differences exhibited already within earlier 
versions \cite{CFG11a}. In the present update, as well as in the preceding 
studies \cite{CFG11,CFG11a}, the steep decrease of the imaginary part of the 
amplitude as function of the nuclear density, due to the single-nucleon nature 
of the model, was identified as a major deficiency of the single-nucleon 
approach. This conclusion is valid also upon adding effective $K^-$-nucleus 
amplitudes generated by the $p$-wave $\Sigma(1385)$ subthreshold resonance, 
as discussed in Ref.~\cite{CFG11a} where $p$-wave effects were found secondary 
to $\Lambda(1405)$-dominated $s$-wave effects. Good agreement with experiment 
was achieved by adding to the potential a phenomenological part which was 
found to be dominated by a $\rho ^2$ or a $\rho ^3$ term. Including 
systematically multi-nucleon processes should be the next step in trying to 
obtain $K^-$-nucleus potentials from in-medium $K^-N$ interaction input.

\section*{Acknowledgements} 
We thank Ale\v{s} Ciepl\'{y} for communicating to us the results of the 
NLO30 chiral model, and him and our other collaborators Daniel Gazda and 
Ji\v{r}\'{i} Mare\v{s} \cite{CFG11,CFG11a} for fruitful discussions. A.G. 
acknowledges gratefully stimulating discussions with Wolfram Weise and the 
hospitality extended to him in Fall 2011 by members of the T39 Group at the 
Technische Universit\"{a}t M\"{u}nchen. This work was supported by the EU 
initiative FP7, HadronPhysics2, under Project No. 227431.

\end{document}